\documentclass[prb]{revtex4}

\usepackage{graphicx}

\begin{document}
\title{Electron-phonon interaction in a superconductor with Kondo scattering}
\author{A.G.Kozorezov$^1$, A. A. Golubov$^2$,
J.K.Wigmore$^1$, D.Martin$^3$, P.Verhoeve$^3$, R.A. Hijmering$^3$}

\address{$^1$ Department of Physics, Lancaster University, Lancaster,
UK,\linebreak $^2$ Department of Applied Physics, University of
Twente, P.O. Box 217, 7500 AE Enschede, The Netherlands,\linebreak
$^3$ Science Payloads and Advanced Concepts Office, SCI-A,
European Space Agency, ESTEC, Noordwijk, The Netherlands}
\date{\today}

\begin{abstract}
In a superconductor with magnetic impurities, Kondo scattering
results in the formation of localized states inside the
superconducting gap. We show that inelastic electronic transitions
involving quasiparticle scattering into and out of the localized
states may result in significant changes in the non equilibrium
properties of the superconductor. Using the model of
Muller-Hartmann and Zittartz for the extreme dilute limit, and
including both deformation potential and spin-lattice coupling we
have calculated the rates of such inelastic transitions between
continuum and discrete states, and shown that they may greatly
modify quasiparticle interactions. The individual processes are:
quasiparticle trapping into discrete states, enhanced
recombination with localized quasiparticles, and pair breaking and
detrapping of localized quasiparticles by sub-gap phonons. We find
that all these processes give rise to clearly distinguishable
temperature dependences of the kinetic parameters.
\end{abstract}

\maketitle \onecolumngrid

\section{Introduction}
The study of the effects of magnetic impurities in superconductors
originated with the pioneering work by Abrikosov and Gor'kov
\cite{Abrikosov}. Recent interest in the subject has been greatly
stimulated by direct observation of the states bound to impurity
atoms \cite{Yazdani} which has led to extensive experimental and
theoretical work in both conventional and unconventional
superconductors\cite{Balatsky}. An important consequence of such
intra gap bound states which has not previously been considered is
their role in providing enhanced trapping and recombination at
impurity atoms in analogy with deep levels in
semiconductors\cite{Poelaert}. Thus, quasiparticles initially in
continuum states may undergo inelastic scattering with phonon
emission and become localized in the vicinity of impurity atoms,
which will act as recombination centers and provide rapid
thermalization of a non equilibrium initial distribution. The
formation of an intra gap band of impurity levels, possibly even
overlapping the ground state, will modify the temperature
dependence of thermalization. Finally, activation of localized
quasiparticles into the continuum spectrum results in an anomalous
temperature dependence of the observable parameters characterizing
the non equilibrium state, such as quasiparticle
lifetime\cite{ourAPL}. Previously neither mechanisms of coupling
nor transitions between the continuum and discrete states bound to
impurities were discussed. In this paper we will show that
electronic transitions between the continuum and bound states
occur both due to deformation potential and spin-lattice
interaction. As will be described later there is strong evidence
that such a scenario has already been observed experimentally.

\quad We will consider the dilute impurity limit $c\ll1$. Here $c$
is the dimensionless impurity density in units of the condensate
density $2N(0)\Delta$, where $N(0)$ is the density of states at
the Fermi level per spin in the normal state, and $\Delta$ is the
gap. The problem can be explicitly solved within the model
originally developed by M$\mathrm{\ddot{u}}$ller-Hartmann and
Zittartz \cite{MHZ} for quantum spins in fully gapped
superconductors. In this model the Hamiltonian of the system is
taken in the form
\begin{eqnarray}\label{Hamiltonian}
H=H_0+H'
\end{eqnarray}
where $H_0$ is the Hamiltonian of an ideal superconductor and $H'$
describes the interaction between impurity atoms and conduction
electrons. The corresponding interaction potential has the
form\cite{Abrikosov}
\begin{eqnarray}\label{potential}
v(\textbf{r})=\sum_i[u_1(\textbf{r}-\textbf{R}_i)+u_2(\textbf{r}-\textbf{R}_i)\sigma\cdot£S]
\end{eqnarray}
where $\textbf{R}_i$ is the coordinate of the impurity atom $S$ is
its spin and $\sigma_{\alpha\beta}^{x,y,z}$ are the spin Pauli
matrices. Here the first term describes the spin independent part
of the impurity scattering potential, and the second term the
exchange interaction. To consider phonon assisted electronic
transitions involving the bound states we must include also terms,
describing the electron-phonon interaction through both the
deformation potential and the spin-lattice interaction; these are
derived by expanding the first and the second terms in expression
(\ref{potential}) respectively to include the displacement of the
impurity atom from its equilibrium site. In the four-dimensional
matrix formalism the full interaction Hamiltonian describing
phonon assisted electronic transitions has the form
\begin{eqnarray}\label{interaction}
H_{int}=\frac{1}{2}\int\mathrm{d}\textbf{r}\psi^+(\textbf{r})V(\textbf{r})\psi^(\textbf{r})
\end{eqnarray}
where $\psi^+(\textbf{r})$ and $\psi(\textbf{r})$ are
four-component operators and $V(\textbf{r})$ is the 4$\times$4
matrix of the form $\left(%
\begin{array}{cc}
  \hat{v} & 0 \\
 0 & -\hat{v}^{tr} \\
\end{array}%
\right)$, where $
\hat{v}(\textbf{r})=\sum_iQ_i\nabla\left(u_1\sigma_0+u_2\sigma_2\cdot£S\right)$,
$\hat{v}^{tr}$ is the transposed matrix, and $Q_i$ is the lattice
displacement of the impurity due to vibrations.

\quad In a superconductor described by the Hamiltonian
(\ref{Hamiltonian}) for an impurity with antiferromagnetic
exchange, bound states split off from the gap are
formed\cite{MHZ}. For the dilute limit the shift of the gap edge
remains small, being proportional to the density of impurities.
Therefore we may disregard all effects of modifications to the
continuum quasiparticle spectrum relating to level shift or to
broadening. Our objective is to determine the spatially averaged
Green function for the continuum spectrum, with the interaction
given by (\ref{interaction}). In the limit $c\ll1$ spatial
averaging can be carried separately for all elements of the Dyson
equation. Indeed, the interaction loop itself will give a
contribution proportional to the number of discrete states and
hence to $c$. Therefore, replacing the external Green functions by
spatially averaged ones introduces inaccuracy in terms only of the
order of $c^2\ll1$, so that the only function remaining to be
averaged is that inside the loop. After spatial averaging, the
Fourier transform to the momentum space, and analytical
continuation from the imaginary to the real axis
$i\omega_n\rightarrow\omega+i\delta$, we may write an expression
for the self energy
\begin{eqnarray}\label{self energy}
&\Sigma_{ph}&(\textbf{p},\omega)=\left(\begin{array}{cc}
 \Sigma_{1,ph}\sigma_0 &  \Sigma_{2,ph}i\sigma_2\\
 -\Sigma_{2,ph}^+i\sigma_2&\tilde{\Sigma}_{1,ph}\sigma_0
\end{array}%
\right)\nonumber\\&=&2\pi\sum_{\textbf{p}'}\sum_{\textbf{q}j}
\int_{-\infty}^\infty\frac{\mathrm{d}z}{2\pi}\int_{-\infty}^\infty\frac{\mathrm{d}z'}{2\pi}\left[\delta(z-\omega_{\textbf{q}j})-\delta(z+\omega_{\textbf{q}j})\right]
\nonumber\\
&\mathrm{Im}&\left(%
\begin{array}{cc}
 G(\textbf{p}',z')|g^+_{\textbf{q},j}(\textbf{p},\textbf{p}')|^2\sigma_0 &  F(\textbf{p}',z') |g^-_{\textbf{q},j}(\textbf{p},\textbf{p}')|^2i\sigma_2\\
 -F^+(\textbf{p}',z') |g^-_{\textbf{q},j}(\textbf{p},\textbf{p}')|^2i\sigma_2& -\tilde{G}(\textbf{p}',z')|g^+_{\textbf{q},j}(\textbf{p},\textbf{p}')|^2\sigma_0 \\
\end{array}%
\right)\nonumber\\&\times&\frac{\tanh\left(\frac{z'}{2T}\right)+\coth\left(\frac{z}{2T}\right)}{\omega-z-z'+i\delta}
\end{eqnarray}
Here $G(\textbf{p}',z')$, $\tilde{G}(\textbf{p}',z')$,
$F(\textbf{p}',z')$ and $F^+(\textbf{p}',z')$ are spatially
averaged electronic Green functions obtained within the model of
M$\mathrm{\ddot{u}}$ller-Hartmann and Zittartz in the dilute
limit. The coupling strength
$g^{\pm}_{\textbf{q},j}(\textbf{p},\textbf{p}')$ is introduced
through
\begin{eqnarray}\label{g}
|g^{\pm}_{\textbf{q},j}(\textbf{p},\textbf{p}')|^2=|g^{(1)}_{\textbf{q},j}(\textbf{p},\textbf{p}')|^2
\pm£S(S+1)|g^{(2)}_{\textbf{q},j}(\textbf{p},\textbf{p}')|^2\nonumber\\
|g^{1,2}_{\textbf{q},j}(\textbf{p},\textbf{p}')|^2=\frac{\hbar}{2MN\omega_{\textbf{q},j}}
(\textbf{p}-\textbf{p}',\textbf{e}_{\textbf{q},j})^2
|u_{1,2}(\textbf{p}-\textbf{p}')|^2
\end{eqnarray}

\quad We obtain for the rate of quasiparticle transitions from the
state ($\textbf{p},\epsilon$) the following expression
\begin{eqnarray}\label{rates}
\Gamma{\textbf{p}}(\epsilon)=-\frac{1}{2\epsilon£Z_1(0)}[(\xi_\textbf{p}+\epsilon)\mathrm{Im}\Sigma_{1,ph}+
(\xi_\textbf{p}-\epsilon)\mathrm{Im}\tilde{\Sigma}_{1,ph}\nonumber\\-\Delta\mathrm{Im}
\left(\Sigma_{2,ph}+\Sigma^{+}_{2,ph}\right)]
\end{eqnarray}
where $Z_1(0)$ is the real part of the renormalization parameter.
Introducing coupling constants analogous to the Eliashberg
constant
\begin{eqnarray}\label{Eliasberg const}
\alpha_{(1,2)}^2(z)F(z)=\frac{\int_{S_F}\frac{\mathrm{d^2}\textbf{p}}{|v_{\textbf{p}}|}\int_{S_F}\frac{\mathrm{d^2}\textbf{p}'}{|v_{\textbf{p}'}|}
\sum_{\textbf{q},j}|g^{(1,2)}_{\textbf{q},j}(\textbf{p},\textbf{p}')|^2\delta(z-\omega_j(\textbf{q}))}{\int_{S_F}\frac{\mathrm{d^2}\textbf{p}}{|v_{\textbf{p}}|}}
\end{eqnarray}
we obtain
\begin{eqnarray}\label{Gamma}
\Gamma(\epsilon)=\frac{\pi}{Z_1(0)}\int\int_{-\infty}^\infty\mathrm{d}z\mathrm{d}z'£F(z)
\{\alpha_{1}^2(z)\mathrm{Re}\left[G_{m}(z')-\frac{\Delta}{\epsilon}F_{m}(z')\right]\nonumber\\
+S(S+1)\alpha_{2}^2(z)\mathrm{Re}\left[G_{m}(z')+\frac{\Delta}{\epsilon}F_{m}(z')\right]\}\nonumber\\
\{\left[\tanh\left(\frac{z'}{2T}\right)+\coth\left(\frac{z}{2T}\right)\right]\delta(\epsilon-z-z')\nonumber\\-
\left[\tanh\left(\frac{z'}{2T}\right)-\coth\left(\frac{z}{2T}\right)\right]\delta(\epsilon+z-z')\}
\end{eqnarray}
Here we have introduced the new notations $G_{m}(z)$ and
$F_{m}(z)$ for the spatially averaged Green functions. The exact
expressions for the $G_{m}(z)$ and $F_{m}(z)$ have the form
\begin{eqnarray}\label{MHZ}
G_{m}(\epsilon)=\frac{\tilde{\epsilon}(\epsilon)}{\sqrt{\tilde{\epsilon}^2(z)-\tilde{\Delta}^2(\epsilon)}};\,\,\,
F_{m}(\epsilon)=\frac{\tilde{\Delta}(\epsilon)}{\sqrt{\tilde{\epsilon}^2(\epsilon)-\tilde{\Delta}^2(\epsilon)}}
\end{eqnarray}
The energy $\tilde{\epsilon}$ and the order parameter
$\tilde{\Delta}$ satisfy the following equations\cite{MHZ}
\begin{eqnarray}\label{MHZ equations}
\tilde{\epsilon}=\epsilon+\Delta\Sigma_1(\tilde{\epsilon},\tilde{\Delta});\,\,\,
\tilde{\Delta}=\Delta+\Delta\Sigma_2(\tilde{\epsilon},\tilde{\Delta})
\end{eqnarray}
where
\begin{eqnarray}\label{MHZ self energies}
\Sigma_1(y,\Delta)=-\frac{c}{i\pi}\frac{\sqrt{y^2-1}}{y^2-y_0^2}y(y-y_0)\nonumber\\
\Sigma_2(y,\Delta)=\frac{c}{i\pi}\frac{\sqrt{y^2-1}}{y^2-y_0^2}y_0(y-y_0)
\end{eqnarray}
Here $y=\epsilon/\Delta$, $y_0=\epsilon_0/\Delta<1$ and
$\epsilon_0$ is the discrete intra gap level. To find solutions to
the main terms, we may use the simplified equations obtained from
(\ref{MHZ equations}) by taking the renormalised parameters
$\tilde{y}=\tilde{\epsilon}/\tilde{\Delta}$ and
$\tilde{y_0}=\epsilon_0/\tilde{\Delta}$ in the denominators in
(\ref{MHZ self energies}). The simplified equations then become
\begin{eqnarray}\label{simplified eqns}
\tilde{z}=z-\bar{c}\frac{y}{|y|}\left(\frac{1}{\tilde{y}-\tilde{y}_0}-\frac{1}{\tilde{y}+\tilde{y}_0}\right)\nonumber\\
\tilde{\Delta}=\Delta+\bar{c}\frac{y}{|y|}\Delta\left(\frac{1}{\tilde{y}-\tilde{y}_0}-\frac{1}{\tilde{y}+\tilde{y}_0}\right)
\end{eqnarray}
where
$\displaystyle{\bar{c}=c\frac{\sqrt{1-y^2}|y|(1-y_0)}{2\pi£y_0}}$.
In order to solve the simplified equations we first note that from
(\ref{MHZ self energies}) $\Sigma_2=-y_0/y\Sigma_1$. Above the gap
edge both $|\Sigma_1|\sim£c$ and
 $|\Sigma_2|\sim£c$; as pointed out above we ignore these corrections.
Inside the gap the solution for the imaginary part $\Sigma''_1$
has the form
\begin{eqnarray}\label{sigma''}
\Sigma''_1=\sqrt{\bar{c}-\frac{1}{4}(|y|-y_0)^2}\Theta\left(\bar{c}-\frac{1}{4}(|y|-y_0)^2\right)
\end{eqnarray}
Within the range $\Sigma''_1\neq£0$, the real part of the
self-energy $\Sigma'_1$ is given by
\begin{eqnarray}
\Sigma'_1=\frac{1}{2}(y_0-|y|)+\frac{\bar{c}}{4y_0}
\end{eqnarray}
Outside this range, but still inside the gap, $\Sigma'_1$ remains
finite with a dependence on impurity concentration changing from
$\sqrt{c}$ at the edge of the $\Sigma''_1\neq£0$ range to $c$ away
from it.

\quad Finally, inside the gap we obtain
\begin{eqnarray}\label{Green intra}
\mathrm{Re}G_{m}(y)=\frac{1+y_0}{(1-y_0)^{3/2}}\sqrt{\bar{c}-\frac{1}{4}(|y|-y_0)^2}\Theta\left(\bar{c}-\frac{1}{4}(|y|-y_0)^2\right)\nonumber\\
\mathrm{Re}F_{m}(y)=\frac{y_0(1+y_0)}{(1-y_0)^{3/2}}\sqrt{\bar{c}-\frac{1}{4}(|y|-y_0)^2}\Theta\left(\bar{c}-\frac{1}{4}(|y|-y_0)^2\right)
\end{eqnarray}
These expressions describe the normalised density of bound states
inside the gap in a superconductor with magnetic impurities. This
distribution is sharp, with both its width and height being
proportional to $\sqrt{c}$. It is easy to confirm that
$\displaystyle{\int_{-1}^1\mathrm{d}y\mathrm{Re}G_{m}(y)=c}$,
corresponding to one quasiparticle bound state for each impurity
atom. Inside the gap, from (\ref{Green intra}), we also obtain
$\displaystyle{ \mathrm{Re}F_{m}(y)=y_0\mathrm{Re}G_{m}(y)}$.

\quad Using the Green functions given by (\ref{Green intra}) and
(\ref{rates}) we may now analyze the different inelastic
transitions. Firstly, a quasiparticle initially in the continuum
state may become trapped. For the trapping rates we obtain
\begin{eqnarray}\label{trap}
\frac{1}{\tau_{trap}(\epsilon)}=\frac{2\Gamma(\omega)}{\hbar}=\int_0^\Delta\frac{\mathrm{d}\epsilon'(\epsilon-\epsilon')}{\Delta^2}\nonumber
\\\left[\mathrm{Re}G_{m}(\epsilon')\left(\frac{1}{\bar{\tau}_{1}}+\frac{1}{\bar{\tau}_{2}}\right)-\frac{\Delta}{\epsilon}\mathrm{Re}F_{m}(\epsilon')
\left(\frac{1}{\bar{\tau}_{1}}-\frac{1}{\bar{\tau}_{2}}\right)\right][n(\epsilon-\epsilon')+1]\nonumber\\{[1-f(\epsilon')]}
=c\left[\frac{1}{\bar{\tau}_{1}}\frac{(\epsilon-\epsilon_0)^2}{\Delta\epsilon}+\frac{1}{\bar{\tau}_{2}}\frac{\epsilon^2-\epsilon_0^2}{\Delta\epsilon}\right][n(\epsilon-\epsilon_0)+1][1-f(\epsilon_0)]
\end{eqnarray}
where $n(\epsilon)$ and $f(\epsilon)$ are the phonon and
quasiparticle distribution functions. The characteristic
relaxation times for phonon assisted scattering on magnetic
impurity in the host lattice $\overline{\tau}_{(1,2)}$  can be
written in the form
\begin{eqnarray}
\frac{1}{\bar{\tau}_{1}}=\frac{1}{\tau_0}\frac{\alpha^2_{1}(\Delta)}{\alpha^2(\Delta)}\left(\frac{\Delta}{T_c}\right)^3;\,\,
\frac{1}{\bar{\tau}_{2}}=\frac{S(S+1)}{\tau_0}\frac{\alpha_{2}^2(\Delta)}{\alpha^2(\Delta)}\left(\frac{\Delta}{T_c}\right)^3
\end{eqnarray}
where $\alpha$ is the parameter entering the Eliashberg constant,
$\tau_0$ is the superconductor characteristic relaxation time for
deformation potential coupling, and $T_c$ is the critical
temperature. The top bars in this notation emphasize that these
characteristic times are for phonon assisted scattering on a
magnetic impurity. An order of magnitude estimate of the ratio
$\tau_0/\bar{\tau}_{1,2}$ can be obtained by direct evaluation of
factors $\alpha_{1,2}^2$. Thus
\begin{eqnarray}\label{couplings}
\frac{\tau_0}{\bar{\tau}_1}\sim£\left(\frac{u_1}{u_{ei}}\right)^2\frac{\hbar£v_s}{\Delta£a_s}\,\,\,\frac{\tau_0}{\bar{\tau}_2}\sim£S(S+1)\left(\frac{u_2}{u_{ei}}\right)^2\frac{\hbar£v_s}{\Delta£a_s}
\end{eqnarray}
where $u_1$, $u_2$ and $u_{ei}$ are the characteristic values of
electron-impurity, exchange interaction and electron-ion
potentials respectively, $v_s$ is sound velocity and $a_s$ is the
radius of the bound state.

 \quad The recombination rate via a bound state calculated
from (\ref{Gamma}) and (\ref{Green intra}) is given by
\begin{eqnarray}\label{recomb}
\Gamma_{R,t}(\epsilon)=\int_0^\Delta\frac{\mathrm{d}\epsilon'(\epsilon+\epsilon')}{\Delta^2}\nonumber
\\\left[\mathrm{Re}G_{m}(\epsilon')\left(\frac{1}{\bar{\tau}_{1}}+\frac{1}{\bar{\tau}_{2}}\right)+
\frac{\Delta}{\epsilon}\mathrm{Re}F_{m}(\epsilon')\left(\frac{1}{\bar{\tau}_{1}}-\frac{1}{\bar{\tau}_{2}}\right)\right]\nonumber\\{[n(\epsilon+\epsilon')+1]}f(\epsilon')
=c\left[\frac{1}{\bar{\tau}_{1}}\frac{(\epsilon+\epsilon_0)^2}{\Delta\epsilon}+\frac{1}{\bar{\tau}_{2}}\frac{\epsilon^2-\epsilon_0^2}{\Delta\epsilon}\right]\nonumber\\{[n(\epsilon+\epsilon_0)+1]}f(\epsilon_0)
\end{eqnarray}
The expression can be written in a more familiar form by
introducing the appropriate recombination coefficient $R^s_t$ and
density of trapped quasiparticles $n_t$ \begin{eqnarray}
\Gamma_{R,t}(\epsilon)=R_tn_t;\,\,\,\, R_t=\frac{1}{2N(0)\Delta}
\left[\frac{1}{\bar{\tau}_{1}}\frac{(\epsilon+\epsilon_0)^2}{\Delta\epsilon}
+\frac{1}{\bar{\tau}_{2}}\frac{\epsilon^2-\epsilon_0^2}{\Delta\epsilon}\right]
\end{eqnarray}
which describes the maximum recombination rate in the absence of a
phonon bottle-neck effect. Comparing the recombination coefficient
on traps $R^s_t$ with that in ideal superconductor $R$ we
obtain\begin{eqnarray}
\frac{R_t}{R}\simeq£\left[\left(\frac{u_1}{u_ei}\right)^2+S(S+1)\left(\frac{u_2}{u_ei}\right)^2\right]\frac{\hbar£v_s}{\Delta£a_s}
\end{eqnarray}
Taking values for
$\left(u_1/u_{ei}\right)^2,\,\,\left(u_2/u_{ei}\right)^2\simeq1$,
$v_s$= 3$\cdot10^5$cm/s, $\Delta=0.5$meV, $S$=1 and $a_s$=1nm we
obtain $R_t/R\geq$10. This ratio indicates the dominance of
recombination at impurities due to the larger magnitude of the
spin-lattice and deformation potential coupling constant with
discrete levels originating from the appearance in (\ref{g}) of a
form-factor for phonon assisted impurity scattering, instead of
the momentum conservation law. Similarly, comparing the maximum
recombination rates under quasi equilibrium conditions for the two
different processes, we obtain
\begin{eqnarray}
\frac{R_tn_{t,T}}{Rn_T}=c\frac{1}{4}\sqrt{\frac{2\Delta}{\pi£T}}\exp\left(\frac{\Delta-\epsilon_0}{T}\right)\nonumber\\
{\left[\left(\frac{\Delta+\epsilon_0}{\Delta}\right)^2
\frac{\alpha_{1}^2}{\alpha^2}+\frac{\Delta^2-\epsilon_0^2}{\Delta^2}
\frac{\alpha_{2}^2}{\alpha^2}\right]}
\end{eqnarray}
where $n_{t,T}$  and $n_T$ are thermal distributions of trapped
and mobile quasiparticles. Hence, even for a small impurity
density, recombination on the traps at low temperatures is a
stronger process because of the presence of the exponential
factor. The presence of this factor significantly accelerates
recombination at low temperatures in superconductors containing
concentrations of magnetic impurities which are below trace
levels. Moreover, the possible formation of an intra gap band of
bound states, and also of discrete bound states in the vicinity of
the Fermi level, can significantly change the observed temperature
dependence of recombination and thermalization rates. In some
situations the rates in an impure superconductor may remain 
finite even at $T=0$.

\quad In spite of the stronger coupling constant the pair breaking
by sub-gap phonons at small impurity density may be less efficient
than for transitions in the continuum spectrum. In the latter case
strong pair-breaking is known to slow down the recombination rate
because of phonon bottle-necking. Thus less efficient phonon
bottle-necking also enhances recombination at impurities. The pair
breaking rate can be calculated using the Green functions
calculated earlier to give
\begin{eqnarray}\label{pair break}
\Gamma_{pb}(\Omega)=\frac{2\tau_0}{\pi\tau_{ph}}
\int_0^{\Omega-\Delta}\frac{\mathrm{d}\epsilon}{\Delta}
\{\left(\frac{1}{\bar{\tau}_1}+\frac{1}{\bar{\tau}_2}\right)\mathrm{Re}G_{m}(\epsilon)\mathrm{Re}G_{m}(\Omega-\epsilon)\nonumber\\+
\left(\frac{1}{\bar{\tau}_1}-\frac{1}{\bar{\tau}_2}\right)\mathrm{Re}F_{m}(\epsilon)\mathrm{Re}F_{m}(\Omega-\epsilon)\}=
c\frac{1}{\tau_{ph}}\eta(\Omega)
\end{eqnarray}
where $\tau_{ph}$ is the characteristic pair breaking time of an
ideal superconductor and
\begin{eqnarray}
\eta(\Omega)=\frac{1}{\pi^2\bar{c}}\left[\frac{\tau_0}{\bar{\tau}_1}+\left(1-2\frac{\epsilon_0}{\Omega}\right)
\frac{\tau_0}{\bar{\tau}_2}\right]\frac{(1+y_0)^{3/2}}{\sqrt{1+\Omega/\Delta-y_0}}\nonumber\\
\int_{\sqrt{\chi(\Omega,y_0)-2\sqrt{\bar{c}}}\Theta(\chi(\Omega,y_0)-2\sqrt{\bar{c}})}^{\sqrt{\chi(\Omega,y_0)
+2\sqrt{\bar{c}}}}\mathrm{d}t\sqrt{4\bar{c}-(\chi(\Omega,y_0)-t^2)^2}
\end{eqnarray}
where $\chi(\Omega,y_0)=\Omega/\Delta-1-y_0$ and $\Theta$ is the
step function.

 \quad The rate of de-trapping from the localized state can be calculated
 without the need for direct evaluation of the broadening of the bound state
 due to spin-lattice interaction. Since we are interested in the de-trapping
 rate due to transitions into all available states, we may balance
 scattering into  and out of the bound states at
 thermal equilibrium. The result is
 \begin{eqnarray}\label{de-trap}
\frac{1}{\tau_{de-trap}}=\frac{1}{\sqrt{2}}\int_{1-\frac{\epsilon_0}{\Delta}}^\infty\mathrm{d}z
\frac{z\exp\left(-z\frac{\Delta}{T}\right)}{\sqrt{z+\frac{\epsilon_0}{\Delta}-1}}\left[\frac{1}{\bar{\tau}_{1}}z+
\frac{1}{\bar{\tau}_{2}}\left(1+\frac{\epsilon_0}{\Delta}\right)\right]
 \end{eqnarray}
The formulas (\ref{trap}), (\ref{recomb}), (\ref{pair break}), and
(\ref{de-trap})
 may be compared with the well known results for an ideal superconductor \cite{Kaplan}.

\quad Experimental data indicating the possible presence of such
processes has appeared previously in several works. Thermalization
which is several orders of magnitude faster than expected is
routinely seen in a number of superconducting absorber materials
used in low temperature single photon detectors. In our own
earlier experiments on non stationary, non equilibrium
quasiparticle distributions in superconducting tunnel junctions
(STJs) we found that many detailed features of the experiments
could only be successfully modelled on the assumption of local
trapping states with well defined energy levels
\cite{acceptedPRB}. However, in the absence of a microscopic model
for the trap, the approach was purely phenomenological. All
inelastic electronic transitions into and out of the traps were
modelled with parameters determined from fitting to experimental
data\cite{acceptedPRB}. In experiments on nominally pure Nb, Ta
and Al STJs and various proximised bi-layer structures, we showed
that the observed behaviour was consistent with the local traps
being either macroscopic regions of suppressed gap as previously
seen in low temperature SEM scans\cite{Gross}, or due to the
presence of magnetic impurities.

\quad Recently the anomalous temperature dependence of
quasiparticle lifetimes in Ta and Al films similar to that
observed in \cite{ourAPL} was reported by R.Barends \emph{et. al}
detected through measurements of kinetic inductance and the
observation of noise spectra\cite{Klapwijk}. An important feature
of the results obtained by this technique was the spatial
homogeneity of the response, indicating the intrinsic character of
the traps and suggesting that they were associated with magnetic
impurities distributed homogeneously throughout the
superconductor. The new mechanisms for inelastic scattering of the
electrons may also be important for electron decoherence in normal
 metals with Kondo impurities, currently the subject of great
 interest\cite{decoherence1, decoherence2}.

\quad We acknowledge valuable discussions with T.M.Klapwijk,
R.Barends and J.R.Gao.
 \onecolumngrid

\end{document}